\documentclass[3p,times]{elsarticle}

\usepackage{color}
\usepackage{amsthm}

\usepackage{amssymb}
\usepackage{amsmath}

\usepackage{lineno}
\usepackage{amsthm}

\biboptions{sort&compress}

\usepackage[figuresright]{rotating}

\begin{document}

\begin{frontmatter}

\title{Critical temperature and upper critical field of  Li$_2$Pd$_{3-x}$Cu$_x$B (x=0.0, 0.1, 0.2) superconductors}

\author{A. A. Castro }
\author{O. Olic\'on}
\author{R. Escamilla}
\author{F. Morales \corref{cor1}}
\ead{fmleal@unam.mx}
\cortext[cor1]{Corresponding author}

\address{Instituto de Investigaciones en Materiales, Universidad Nacional Aut\'onoma de M\'exico. Ciudad de M\'exico, 04510, M\'EXICO.}

\begin{abstract}
We studied the effects of substitution of Pd by Cu on the upper critical field of the noncentrosymmetric superconductor  Li$_2$Pd$_{3-x}$Cu$_x$B, with x=0.0, 0.1 and 0.2. The upper critical field as a function of temperature was determined by resistance measurements  at different magnetic fields. We found that the superconducting transition temperature decreases as the Cu content increases. Moreover, the temperature dependence of the upper critical field is linear in the range of the temperature studied and, at low temperature, is enhanced compared with the prediction of the Werthamer-Helfan-Hohenberg theory. This indicates that the breaking of Cooper pairs by spin orbit scattering and Pauli paramagnetism is negligible, and that the upper critical field enhancement is mainly because the electron-phonon coupling and disorder.
\end{abstract}


\end{frontmatter}

%\linenumbers
% \linenumbers

\section{Introduction}

Noncentrosymmetric superconductors  present a nonstandard superconducting behavior. The absence of inversion symmetry produces an antisymmetric spin-orbit coupling (ASOC) that modify the electronic behavior, and   Cooper pairs tend to pair in a mixture of spin-singlet and spin-triplet symmetry depending of the ASOC intensity, producing nodes or lines of nodes in the energy gap function \cite{bauer,Nishiyama2005,Yokoya2005,Nishiyama2007}.

The noncentrosymmetric superconductor Li$_2$Pd$_3$B is considered a conventional superconductor without strong electronic correlations \cite{Nishiyama2005,Yokoya2005,Mani2009}, with  transition temperature about $T_C \sim$ 7-8 K. The substitution of Pd with Pt induces important changes, for example, the $T_C$ value decreases to 2.2 K and the system displays a mixture of spin--singlet and spin--triplet pairing; thus the superconducting energy gap function  has lines of nodes \cite{Nishiyama2007,Takeya2007,Harada2012}.

 Substitution of elements in Li$_2$Pd$_3$B modifies its electronic characteristics, particularly the substitution of Pd renders modifications of the electronic density of states (DOS) and in consequence the superconducting properties, because the Pd 4$d$ electrons are considered responsible of superconductivity \cite{Mani2005,Lee2005,Mani2009}. Partial substitution of Pd with Ni, produces a decrement of $T_C$ and upper critical field at $T=0$ K ($H_{C2}(0)$) as the Ni content increases \cite{Mani2005,Li2008,Mani2009}. Substitution with a nonmagnetic element, as Al in the B sites, causes a decrement of $T_C$ but $H_{C2}(0)$ becomes increased by a factor of 1.5 \cite{Bao2012}. It is noteworthy that in both substitutions  $H_{C2}(T)$ shows a linear behavior in the temperature and in the impurity range studied. The linear behavior of $H_{C2}(T)$ has been observed in two band superconductors \cite{Kim2007,Bay2012,Hunte2008} and in isotropic gaped superconductors  \cite{kim2006}.

The Werthamer-Helfand-Hohenberg (WHH) theory \cite{Helfand1966,Werthamer1966} provides a way to study the effects of
the spin paramagnetism and the spin-orbit scattering on the
upper critical field behavior of superconducting materials. This theory takes into account the Cooper pair breaking with the Pauli paramagnetism and the spin-orbit scattering. The presence of these processes produce a decrement of $H_{C2}(T)$ at low temperatures, as compared with the upper critical field when they are not present, and it is linear near $T_C$.

Until now, a study of the effect of substitution of Pd with Cu, a nonmagnetic element, has not be reported yet.
In this paper we report the effects of partial substitution of Pd with Cu on the critical temperature and on the temperature dependance of the upper critical magnetic field of Li$_2$Pd$_{3-x}$Cu$_x$B with x=0.0, 0.1 and 0.2.
The results show that $T_C$ decreases with the Cu increment faster than when Pd is substituted with a magnetic element. Forward, $H_{C2}(T)$ shows a linear behavior and enhances at low temperatures deviating from the WHH theory. This behavior indicates that the spin-orbit scattering and the Pauli paramagnetism take a minor role in the Cooper pairs breaking.

\section{Experimental details}

 Polycrystalline samples of Li$_2$Pd$_{3-x}$Cu$_x$B with x=0.0, 0.1 and 0.2, were synthesized by arc-melting using Li (99+\%), Pd (99.95\%), Cu (99.99\%) and B (99.99\%) powders as precursors.
 The synthesis was performed in argon atmosphere, following the two-step process as reported by Togano et al. \cite{Togano2004} and  adding 20\%  Li excess to compensate losses.
 The obtained ingots were processed  several times in order to get homogeneous samples. Sample structure was determined by X-Ray diffraction,  X-ray patterns of the powdered samples were acquired at room temperature using a Siemens D-5000 diffractometer with Co-K$_\alpha$ radiation ($\lambda=1.79026$ \AA) and Fe filter. Patterns were obtained in steps of 0.015$^\circ$ at  8 s in the 2$\theta$ range of 20$^\circ$  to 100$^\circ$. Rietveld analysis of diffraction patterns was performed with the MAUD program \cite{maud}.

 Electrical resistance versus temperature ($R(T)$) and magnetic field was measured using the four-probe technique in a Physical Properties Measurement System (Quantum Design). $R(T)$ measurements, without applied magnetic field, were measured from room temperature to 2 K. The magnetoresistance measurements were performed between 2 K and 10 K with applied magnetic field between 0 and 40 kOe.

\begin{figure}[t]
\begin{center}
\includegraphics[scale=0.28]{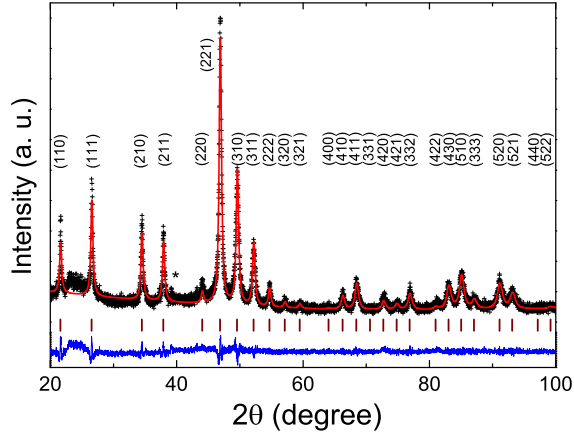}
\caption{\label{rx}(Color online) X-ray patterns of Li$_2$Pd$_{3-x}$B and  Rietveld refinement (continuous line). Vertical lines are the reflections,  reflection at 39.3$^\circ$ corresponds to a tiny amount of Pd$_2$B$_5$ impurity. The line at the bottom is the difference between experimental and refined pattern. }
\end{center}
\end{figure}

\section{Results and discussion}

As mentioned, the Li$_2$Pd$_{3}$Cu$_x$B samples were analyzed by X-ray powder  and  diffraction patterns fitted by Rietveld method, Fig. \ref{rx} shows the refined pattern of Li$_2$Pd$_{3-x}$B sample. Vertical lines, at the bottom of the panel, are the reported reflections (ICSD 84931), where  the Miller indexes of each plane are  indicated. Tiny amount of impurities of Pd$_2$B$_5$ was identified in the sample, it is indicated with *. The difference between the experimental pattern and the refined pattern is show at the bottom of the figure. Refinements were performed using the space group $P4_332$ as used in previous reports \cite{Eibenstein1997,Mani2005,Bao2013}. Lattice parameters of the cubic structure obtained from the Rietveld analysis  are: 6.7427(2) \AA, 6.7363(2) \AA\ and 6.7401(2) \AA\ for x=0, 0.1 and 0.2, respectively. The value obtained for the sample without Cu is in agreement to reported values \cite{Togano2004,Badica2005,Mani2005,Eibenstein1997}.

\begin{figure}[h!]
\begin{center}
\includegraphics[scale=0.28]{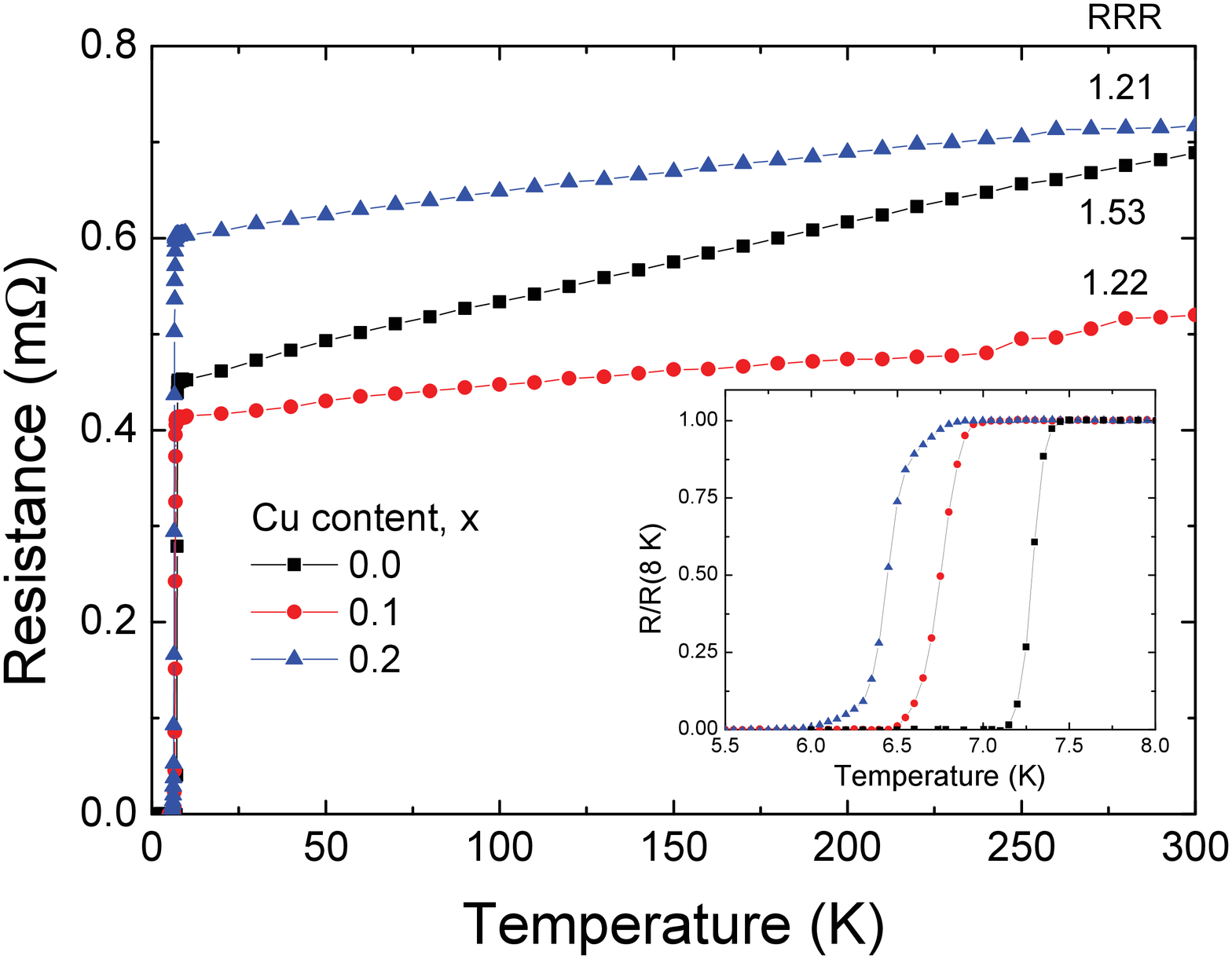}
\caption{\label{rt2}(Color online) Electrical resistance as a function of temperature of Li$_2$Pd$_{3-x}$Cu$_x$B samples. The residual resistance ratio ($RRR$) is indicated for each sample. The inset shows the low temperature normalized resistance at 8 K, there the superconducting transition is observed.}
\end{center}
\end{figure}

The electrical resistance as a function of temperature $R(T)$, between low and room temperature, of the studied samples is shown in Fig. \ref{rt2}. There, the residual resistance ratio ($RRR=R_{300 K}/R_{8 K}$) is indicated. The $RRR$ value provides qualitative information about electron scattering by impurities and vacancies. The $RRR$ values indicate that our samples have a big number of defects, probably included Li vacancies, because the low melting point of Li, and possibly substitutional Cu defects. Similar values of $RRR$ were reported in Li$_2$Pd$_3$B polycrystalline samples with analogous $T_C$ \cite{Mani2009,Badica2005}, however samples of better quality, with $RRR \sim$ 6.5 and $T_C \simeq 8$ K, has been reported  \cite{Togano2004}.

\begin{figure}[t]
\begin{center}
\includegraphics[scale=0.35]{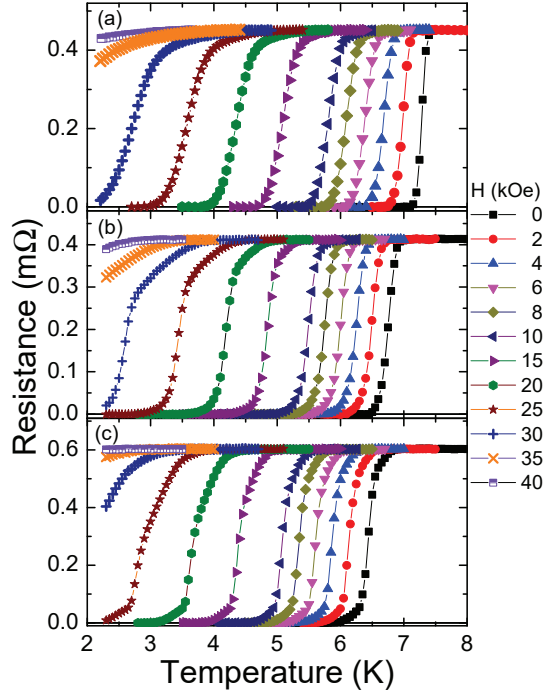}
\caption{\label{rth} (Color online) Electrical resistance as a function of temperature and magnetic field for Li$_2$Pd$_{3-x}$Cu$_x$B, (a) x=0.0, (b) x=0.1 and (c) x=0.2. The symbols identify the applied magnetic field and are the same for the three panels.}
\end{center}
\end{figure}

The electrical resistance as a function of temperature measurements of Li$_2$Pd$_{3-x}$Cu$_x$B shows that $T_C$ decreases as the Cu content increases. The inset in Fig. \ref{rt2} shows the normalized resistance at 8 K, it is clear the Cu effect on the $T_C$. The critical temperature was determined at the middle of the transition.
$T_C$ values were  7.29 K, 6.75 K, and 6.44 K for samples with x=0.0, 0.1, and 0.2 respectively. The $T_C$ value of the sample without Cu is between the values reported in the literature \cite{Bao2012,Badica2004,Arima2013}. The transition temperature of Li$_2$Pd$_{3-x}$Cu$_x$B samples is sharp, with the width of the transition temperature $\Delta T$ values between 0.16 and 0.45 K, $\Delta T$ increases as the Cu content increases. A coarse estimation of the slope of $T_C(x)$ is about -4.25 K/x, where x is the nominal at\% Cu into the samples. Determining the slope of $T_C(x)$ from the Li$_2$(Pd$_{1-x}$Ni$_x$)$_3$B  \cite{Mani2005} and Li$_2$(Pd$_{1-x}$Pt$_x$)$_3$B \cite{Badica2005} data (in these cases x is the Ni or Pt content, respectively), we obtained -2 K/x and -5 K/x, respectively. From these values, it is clear that substitution of Pd with Cu or Pt produces a stronger decrement of $T_C$ than the substitution of Pd with Ni. Then, a magnetic impurity into Li$_2$Pd$_3$B produces a weak decrement of $T_C$ than a non magnetic impurity. This fact disagrees with the idea that Li$_2$Pd$_3$B is a conventional superconductor, where the magnetic impurities must have a strong effect in $T_C$.

The magnetoresistance measurements are shown in Fig. \ref{rth} for the three studied samples with  applied fields from 0 to 40 kOe, as expected the critical temperature decreases as the magnetic field is increased. In the temperature range studied the lower $T_C$ was determined at $H=30$ kOe, for higher fields  $T_C$ was not possible to determine because the temperature limit of the apparatus.

\begin{figure}[t]
\begin{center}
\includegraphics[scale=0.3]{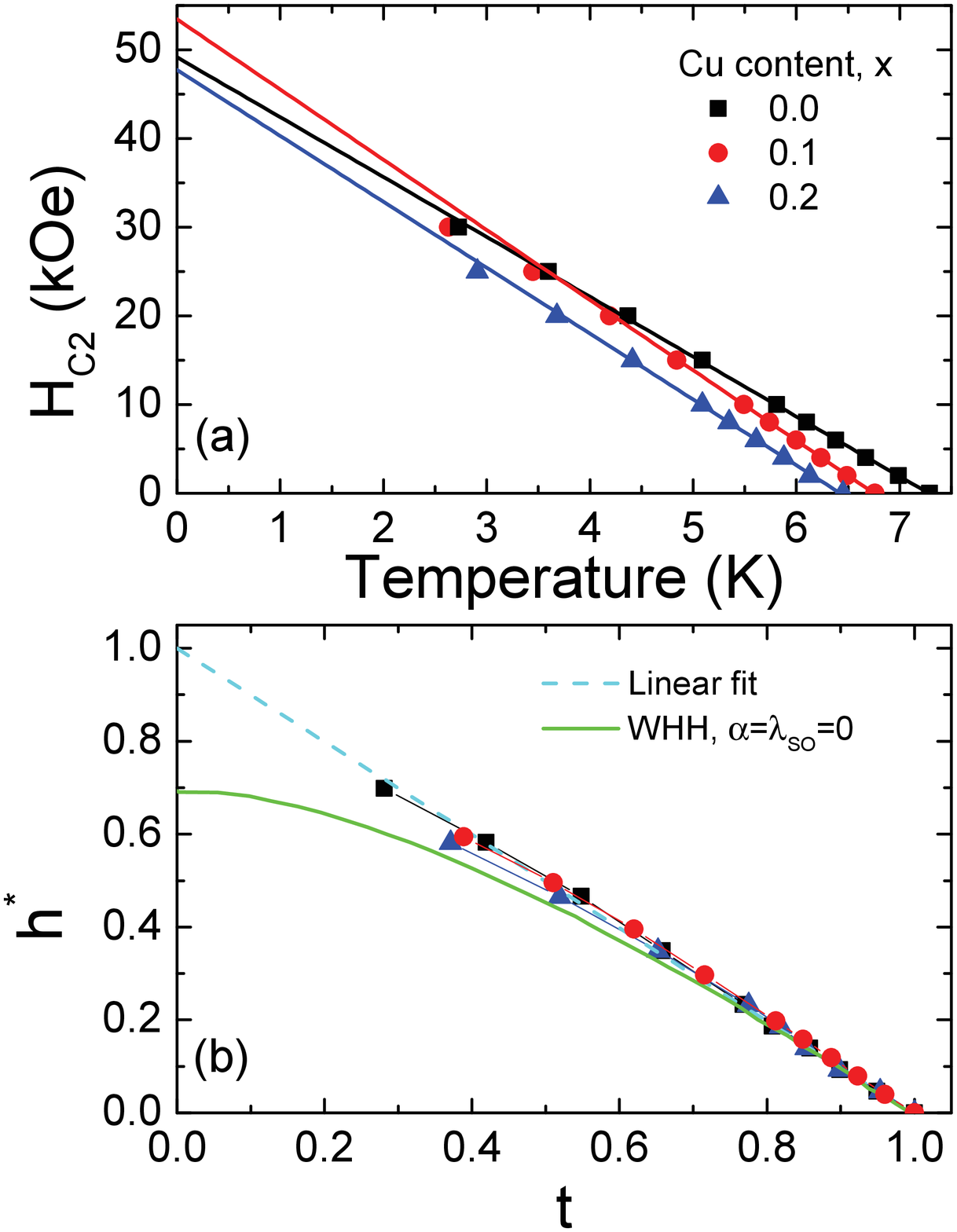}
\caption{\label{hc2t}(Color online) (a) Upper critical field as a function of temperature of Li$_2$Pd$_{3-x}$Cu$_x$B (x=0.0, 0.1 and 0.2). The lines are a linear fit of the data near to $T_C$ (b) Normalized upper critical field $h^*$ as a function of normalized temperature $t=T/T_C$. Continuous line represents the WHH model calculated with $\alpha=\lambda_{SO}=0$. The dashed line is a linear fit of data near $t=1$ extrapolated to $t=0$. }
\end{center}
\end{figure}

The upper critical field as a function of temperature was extracted from the curves of Fig. \ref{rth}. $H_{C2}(T)$ shows a linear behavior in the temperature range measured, see  Fig. \ref{hc2t}(a). The linear extrapolation at T=0 K gives values of $H_{C2}(0)$ between 48 to 54 kOe, with the higher value for sample with x=0.1. Figure \ref{hc2t}(b) shows the reduced upper critical field $h^*$ versus the reduced temperature $t=T/T_C$, for the three studied samples. To plot these data we used the  equation proposed by WHH \cite{Helfand1966,Werthamer1966}:
\begin{equation}
h^*=\frac{H_{C2}(t)}{\left(-\frac{dH_{C2}}{dt}\right)_{t=1}}.
\end{equation}

It is noteworthy the linear behavior of the experimental data between $t=1$ and $t\simeq 0.3$ and those almost collapse in one line. Similar linear behavior was observed in Li$_2$(Pd$_1{-x}$Ni$_x$)$_3$B but in a reduced temperature range, from $t=1$ to $t\simeq 0.6$ \cite{Mani2009}. To get information about the Cooper pairs breaking, we compare the experimental data with the WHH theory (continuous line Fig. \ref{hc2t}(b)). This curve was calculated according with \cite{Werthamer1966},
\begin{equation}
%\begin{split}
ln\frac{1}{t}  =  \sum_{\nu=-\infty}^{\infty}
\left\{ \frac{1}{|2\nu+1|}
 - \left [|2\nu+1| + \frac{\bar{h}}{t}
 + \frac{(\alpha \bar{h} /t)^2}{|2\nu+1| + (\bar{h}+\lambda_{SO})/t} \right]^{-1} \right \},
%\end{split}
\end{equation}
in this equation $\bar{h}$ is a dimensionless upper critical magnetic field, $\alpha$ is the Maki parameter and $\lambda_{SO}$ is the spin-orbit scattering, $\alpha$ and $\lambda_{SO}$ were taken as zero. It is clear that the experimental data deviate from the WHH prediction. Under this condition the spin-orbit scattering and pair breaking due to Pauli paramagnetism may be considered negligible.

The upper critical field at 0 K of the samples was determined in the approximation proposed in the WHH theory \cite{Helfand1966,Werthamer1966};
\begin{equation}
H_{C2}^{WHH}(0)=-0.693 T_C \left(\frac{dH_{C2}}{dT}\right)_{T=T_C}
\label{h0}
\end{equation}

The $H_{C2}^{WHH}(0)$ values obtained for Li$_2$Pd$_{3-x}$Cu$_x$B are shown in Table \ref{whh}. These values are slightly lower than the values reported for Li$_2$Pd$_3$B \cite{Togano2004,Landau2007,Doria2006}. Samples where Pd was partially replaced with Ni shown that $H_{C2}(0)$ and $T_C$ decrease as the Ni content increases. The $H_{C2}(T)$ of these samples is almost linear \cite{Mani2009}.
Substitution of a nonmagnetic impurity, to produce Li$_2$Pd$_3$B$_{1-x}$Al$_x$ samples (x=0.0-0.1), shown small $T_C$ decrements and an increment of $H_{C2}(0)$, until 1.5 times. The $T_C$ as a function of Al content or residual resistivity does not show a tendency. The upper critical field increment is attributed to impurities and defects, based on the residual resistivity, in fact the sample with the poor quality shown the higher $H_{C2}(0)$ and correspond to the sample without Al \cite{Bao2012}.

\begin{table}[t]
\begin{center}
\caption{\label{whh} Superconducting critical temperature $T_C$, slope of $H_{C2}(T)$ at $T=T_C$ and upper critical field $H_{C2}$ at T=0 K determined by the WHH approximation for Li$_2$Pd$_{3-x}$Cu$_x$B samples. The coherent length at $T=0$, $\xi_0$, is included. }
\begin{tabular}{ccccc}
\hline
Cu content & $T_C$  & $-(\frac{dH_{C2}}{dT})_{T=T_C}$  & $ H_{C2}^{WHH}(0)$ & $\xi_0$  \\
x   &    (K)    &   (kOe/K)    &    (kOe) & (nm)    \\
\hline
0.0   & 7.29   & 6.75  & 34.1 & 9.84 \\
0.1 & 6.75   & 7.92  & 37.0 & 8.91 \\
0.2 & 6.44   & 7.43  & 33.1  & 9.96 \\
\hline
\end{tabular}
\end{center}
\end{table}

The Ginzburg-Landau equation,  $H_{C2}(0)= \Phi_0 (2\pi \xi_0^2)^{-1}$, relates the upper critical field with the coherent length $\xi_0$. In this equation $\Phi_0=h/2e=2.078\times10^{-15}$ T m$^2$ is the quantum fluxoid.
Using the $H_{C2}^{WHH}(0)$ values, we estimated $\xi_0$, the results where included in Table \ref{whh}. Similar values were reported previously for Li$_2$Pd$_3$B \cite{Takeya2009,Bao2012,Bao2013}.

Two non expected behavior has been observed in the Li$_2$Pd$_{3-x}$Cu$_x$B samples. The firs one is a higher decrement of $T_C$ with the Cu content, as compared with Li$_2$Pd$_3$B where Pd was substituted by Ni, a magnetic element. The second unexpected behavior is an enhancement of the $H_{C2}(T)$ at low temperatures that deviates from the WHH theory. This enhancement has been related to different causes; strong electron-phonon coupling \cite{Bulaevskii1988}, anisotropic Fermi surface \cite{Kita2004} and localization effects in highly disordered superconductors \cite{Coffey1985}. The intensity of the electron-phonon coupling can be known from the BCS ratio $2\Delta/K_BT_C$=3.52, indicative of weak coupling, or from the electron-phonon coupling constant $\lambda_{e-ph}$. $2\Delta/K_BT_C$ values between 3.94 and 4.5  \cite{Takeya2005a,Hafliger2009,Tsuda2009} and $\lambda_{e-ph}$ values between 0.74 and 1.09 \cite{Bose2005,Mani2009,Takeya2005a} have been reported for Li$_2$Pd$_3$B. These values indicate that Li$_2$Pd$_3$B is an intermediate coupling superconductor. Then the $H_{C2}(T)$ enhancement may be because this characteristic. However some grade of disorder could be present due to a high number of defects, as indicated by the RRR values, then disorder could be participating in the enhancement of $H_{C2}(T)$. It is clear that more work is required to explain this behavior, particularly it is necessary to know the possible role of Li vacancies in this problem.

\section{Conclusions}
In summary, we reported the effects of partial substitution of Pd with Cu in Li$_2$Pd$_3$B on the superconducting critical temperature and on the upper critical field. The results shown that $T_C$ decreases as the Cu content is increased, with a slope of $T_C$(x) of -4.25 K/x. The decrement of $T_C$ with Cu content is faster than the decrement produced by a magnetic impurity, suggesting that Li$_2$Pd$_3$B may be considered an unconventional superconductor. The upper critical field, at low temperature, is enhanced as compared with the WHH theory, without Pauli paramagnetism and spin-orbit scattering pair breaking contribution.

\section*{Acknowledgments}
We thank R. Escudero for useful discussion and critical reading of the manuscript.

\section*{References}

\bibliographystyle{elsarticle-num}
\bibliography{lipdb}

\end{document}